*metals*

*Article*

# Effect of Substrate Roughness on Oxidation Resistance of an Aluminized Ni-base Superalloy

**Wojciech J. Nowak \*, Kamil Ochał, Patrycja Wierzba, Kamil Gancarczyk and Bartek Wierzba**

Department of Materials Science, Faculty of Mechanical Engineering and Aeronautics, Rzeszow University of Technology, Powstancow Warszawy 12, 35-959 Rzeszow, Poland; kochal@prz.edu.pl (K.O.); PatrycjaWierzba@prz.edu.pl (P.W.); KamilGancarczyk@prz.edu.pl (K.G.); bwierzba@prz.edu.pl (B.W.)

\* Correspondence: w.nowak@prz.edu.pl; Tel.: +48-17-743-2375

**Abstract:** In the present work, it is shown that the surface preparation method used on two Ni-based superalloys prior to aluminizing chemical vapor deposition (CVD) is one of the most important factors determining the oxidation resistance of aluminized Ni-based superalloys. It was found that grit-blasting the substrate surface negatively affects the oxidation resistance of the aluminized coatings. For grit-blasted and aluminized IN 625, a thicker outer NiAl coating was formed compared to that of IN 738. In contrast, no effect on NiAl coating thickness was found for grit-blasted and aluminized IN 738. However, a thicker interdiffusion zone (IDZ) was observed. It was shown that the systems with grit-blasted surfaces reveal worse oxidation resistance during thermal shock tests—namely, a higher mass loss was observed for both grit-blasted and aluminized alloys, as compared to ground and aluminized alloys. A possible reason for this effect of remaining alumina particles originating from surface grit-blasting on the diffusion processes and stress distribution at the coating/substrate is proposed.

## 1. Introduction

Modern materials used in the hottest parts of stationary gas turbines or aircraft engines are facing very aggressive environments at high temperatures. This leads to the oxidation of the base materials, which, in turn, causes oxide scale formation and spallation, resulting in component walls' thickness decrease and the losing of mechanical properties at elevated temperatures [1]. Moreover, the temperature at the hottest parts of gas turbines causes the rapid corrosion of used materials such as Ni-based superalloys. Therefore, further protection of the components needs to be provided [2]. One of the methods to increase the oxidation resistance of Ni-base superalloys is the production of protective coatings such as MCrAlY-type coatings [3] or aluminide layers [4]. Protective aluminide coatings can be produced by in-pack cementation [5,6], physical vapor deposition or chemical vapor deposition [7], using slurries containing Al [8,9], or additive manufacturing [10,11]. Though the oxidation behavior of aluminide coatings on Ni-based superalloys has been widely studied [12–14], the effect of a substrate surface preparation method on oxidation behavior of aluminide layers on Ni-based superalloys has been rarely studied. Recently, Chen et al. [15] studied the effect of substrate surface sand-blasting prior to coating of a glass matrix composite of an Ni-base superalloy on its oxidation behavior at 1000 °C in air. The authors found that the system with sand-blasted surface exhibited worse oxidation resistance as compared with the same system with a polished substrate surface prior coating. Similar research was conducted by Dong et al. [16]. The authors investigated the effect of substrate roughness and aluminizing agent composition on the coating surface roughness, phase structure, and the thickness of each phase present in an aluminide coating formed in the coating of a stainless steel by the pack cementation method. It was observed that the different substrate surface preparation methods resulted in a different coating surface roughness. It has recently been shown that surface preparation methods, namely polishing, grinding and grit-blasting, detrimentally influence the oxidation behavior of Fe-based alloys [17], Ni-base superalloys [18], and even pure metallic elements, e.g., Cu, Ni, and their alloys [19]. Recently, Ramsay et al. [20] observed the negative influence of the introduced stress on the oxidation behavior of the Ni-base superalloy RR1000. One can suspect that grit-blasting increases stresses in the near-surface region. Therefore, an effect on oxidation behavior can be expected. Sun et al. [21] used electrolytic polishing to decrease the roughness of an aluminide layer on an Fe-base substrate during studies on formation and phase transformation of an aluminide coating prepared by a low temperature aluminizing process. However, no research regarding the effect of the substrate surface preparation method on Ni-base superalloys' aluminizing process, as well as the oxidation behavior of such aluminide systems, has been performed. Therefore, the aim of the present study was to examine the role of Ni-base surface mechanical treatment on the microstructure and oxidation behavior of aluminide coatings produced by high temperature low-activity chemical vapor deposition on Ni-base superalloys during exposure at a high temperature.

## 2. Materials and Methods

In this work, aluminide coatings produced on two commercially available Ni-base superalloys from the Inconel family—namely IN 625 and IN 738—with the nominal composition given in Table 1 were investigated in terms of their oxidation resistance under thermal shock conditions. Prior to aluminizing, all samples were ground using 220 grit sand paper. Then, one of the samples from each alloy was ground using sand paper with an increasing gradation up to 1000 grit. Other samples were grit-blasted using an alumina powder with grain dimensions of approximately 60 μm (220 mesh). The grit-blasting pressure was 0.8 MPa, and the nozzle diameter was 1.5 mm. After preparation, all of the samples were cleaned ultrasonically in ethanol and dried by compressed air. The surface roughness of all samples was measured using a HOMMEL Werke T8000 (Hommelwerke GmbH, VS-Schwenningen, Germany) contact profilometer. Substrate surface topography was reproduced using a Sensofar S-Neox Non-contact 3D Optical Profiler (Sensofar, Barcelona, Spainwith a vertical resolution of 1 nm (Figure 1a,b). It should be mentioned that the *Y*-axis was higher for a grit-blasted surface compared to a ground surface. The samples with ground surfaces revealed anisotropic roughness, i.e., the grinding direction could be clearly observed (Figure 1a). To exclude the effect of any anisotropy of the ground surface, the roughness measurement was always performed in the direction perpendicular to the grinding direction. In contrast, the surfaces prepared by grit-blasting revealed isotropic roughness (Figure 1b). After roughness measurements, all of the samples were subjected to an aluminizing process by the low-activity chemical vapor deposition (CVD) method using the BPXPRO3242 equipment of the IonBond company in the R&D Laboratory for Aerospace Materials, Rzeszów University of Technology, Poland. The gaseous atmosphere consisting of a mixture of $AlCl_3$ and $H_2$ was produced in an external generator and transferred into a retort where samples had been placed. The process was performed at 1040 °C for 6 h. After aluminizing, parts of the samples designated for characterization in the as-received condition stage were subjected to phase analysis using a Miniflex II X-ray diffractometer made by Rigaku ( Tokyo, Japan). For the X-ray source, a filtered copper lamp ($CuK\alpha, \lambda = 0,1542$ nm) with a voltage of 40 kV was used. The $2\theta$ angle range varied between 20 and 120°, and the step size was 0.02°/s. Phase composition was determined by using the Powder Diffraction File (PDF) developed and issued by the ICDD (The International Center for Diffraction Data). After the XRD measurement, the samples were investigated using a glow discharge optical emission spectrometer (GD-OES) made by Horiba Jobin Yvon (Longjumeau, France ). The GD-OES depth profiles were quantified using the procedure described in references [22–24]. Another part of the samples was investigated in terms of their oxidation resistance under thermal shock conditions. The thermal shock test was carried out in furnace SCZ 120/150 made by Czylok (Jastrzębie Zdrój, Poland ). To shorten the time of the test, an extreme thermal condition was applied, i.e., the test was carried out at 1120 °C under a cyclic oxidation loop consisting of 2 h of heating and 15 min of cooling with pressurized air. When the furnace was heated to 1120 °C, the sample holder was in the cooling position outside the furnace. After reaching a pre-set temperature, the arm with the samples was automatically moved into the hot zone of the furnace and oxidized for 2 hours. Considering relatively thin samples (about 3 mm in total), the incubation stage (time to reach 1120 °C by the samples) was about two to three minutes and could be neglected. After 2 hours of heating, the arm with the samples was automatically moved into the cooling zone and cooled with pressurized air to room temperature. The time to reach room temperature by the samples was maximally 3 minutes. Therefore, the heating and cooling rate were severe. An inspection for measuring the weight change of the samples and a visual check of the condition of their surfaces were carried out every 10 cycles. The thermal shock test was performed up to 200 cycles (400 hot hours). The cross-sections of the samples in the as-received conditions and after exposure were prepared in the following way: After exposure, the samples were sputtered with a thin gold layer, subsequently electroplated with nickel, and then mounted in epoxy resin. The fine polishing with an $SiO_2$ suspension with 0.25 μm granulation was the final step in the preparation of metallographic cross-sections. The cross-sections were analyzed by a light optical microscope (LOM) Nikon EPIPHOT 300 (Nikon, Tokyo, Japan) and scanned by using a scanning electron microscope (SEM), Hitachi S3400N (Hitahi, Tokyo, Japan), equipped with an energy dispersive spectroscopy (EDS) detector. The SEM analysis was performed to investigate the surfaces of the exposed samples prior cross-sectioning. The thickness of the coatings presented in Table 2 were calculated based on 13 measurements on three different, randomly chosen locations on the cross-sections.

**Table 1.** Nominal chemical composition of Ni-base alloys used as a substrates.

| Alloy | Elements Content (wt %) | | | | | | | | | | | | | |
|-------|------|-------|--------|------|------|-----|--------|------|------|------|------|-----|------|------|
|       | Ni   | Cr    | Ta     | Co   | Mo   | W   | Nb     | Al   | Ti   | Fe   | Mn   | Si  | B    | C    |
| IN 625 | BASE | 21.50 | 3.65 * | 1.00 | 9.00 | -   | 3.65 * | 0.40 | 0.40 | 5.00 | 0.5  | 0.5 | -    | 0.10 |
| IN 738 | BASE | 16.00 | 1.8    | 8.5  | 1.8  | 2.6 | 0.8    | 3.5  | 3.5  | -    | -    | -   | 0.01 | 0.18 |

\* Total content of Ta and Nb.

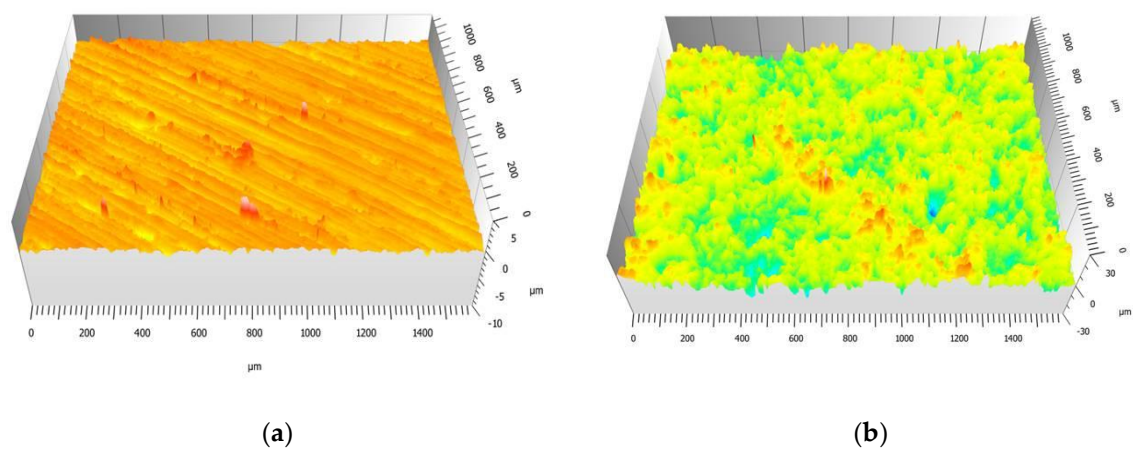

(**a**)                                                            (**b**)

**Figure 1.** Reproduction of the: (**a**) Ground and (**b**) grit-blasted surfaces topography of the IN 625 substrate prior to the chemical vapor deposition (CVD) aluminizing process.

## 3. Results and Discussion

### 3.1. Roughness Evaluation of Ni-Base Substrates Surfaces

Figure 2 shows the exemplary results of the measured surface roughness profiles obtained on the surface of IN 625 before the CVD process. One should notice that the *Y*-axis was 10x higher for grit-blasted surface than for the ground surface. Based on obtained roughness profiles, roughness parameters $R_a$ were calculated for the ground surface $R_a$ = 0.253 μm. For the grit-blasted surface, $R_a$ =

3.248 µm. The $R_a$ values measured for IN 738 were very similar. Therefore, they are not shown here. One can notice that the difference in the $R_a$ parameter values was about one order of magnitude higher for the grit-blasted surface as compared to the ground one. This observation is in agreement with the literature data [17,18].

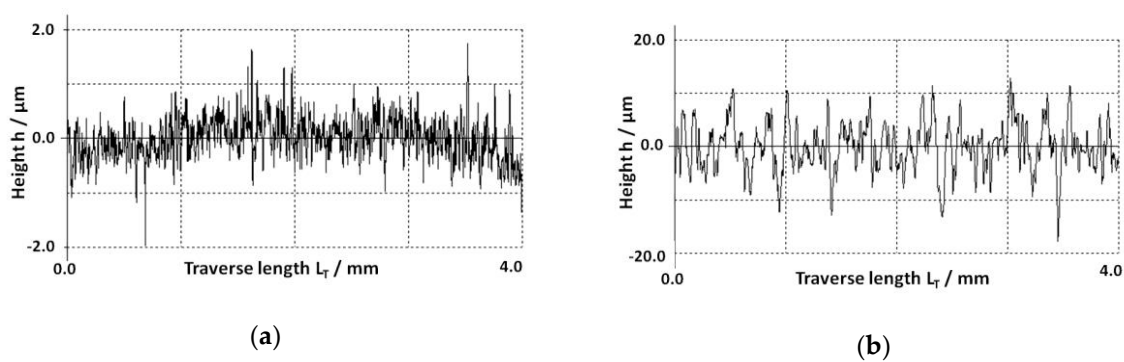

(a)                                                        (b)

**Figure 2.** Surface roughness profiles of IN 625 performed by standard contact HOMMEL Werk T8000 profilometer on: (**a**) Ground (1000 grit) and (**b**) grit-blasted Ni-base superalloy.

### 3.2. Aluminide Coatings in the As-Received Condition

To investigate the effect of substrate roughness on the roughness of an aluminide coating, a surface roughness measurement of the aluminide coatings in the as-received condition was carried out using a standard contact profilometer. The results of these measurements are summarized in Figure 3. As shown in Figure 3, in the case of aluminide coatings formed on ground substrates, an increase of surface roughness measured on aluminide coatings in comparison with substrate surface roughness was observed. However, it should be mentioned that the $R_a$ value measured on the aluminide coating manufactured on ground IN 625 was about three times higher compared to the $R_a$ value for the ground substrate, while the $R_a$ value measured on aluminide layer produced on IN 738 increased less than two times. In contrast, the manufacturing of aluminide coating by CVD on grit-blasted substrates resulted in a decrease in the $R_a$ value for the aluminide coatings in comparison with the grit-blasted substrate. Moreover, it was observed that the aluminide coating on IN 738 was rougher than the one deposited on IN-625. It is known that in the case of a low activity chemical vapor deposition (LA-CVD), an aluminide layer grows outward via the transport of Ni and a reaction with Al in the gaseous phase on the surface of the substrates [25]. Therefore, the changes between the roughness of substrates before the CVD aluminizing process and aluminide coatings were probably the consequence of the mass transport processes during the formation of aluminide coatings. Despite different trends of roughness changes, one can clearly see that the surface roughness obtained for aluminide layers formed on the grit-blasted substrates were higher than for these formed on the ground surfaces.

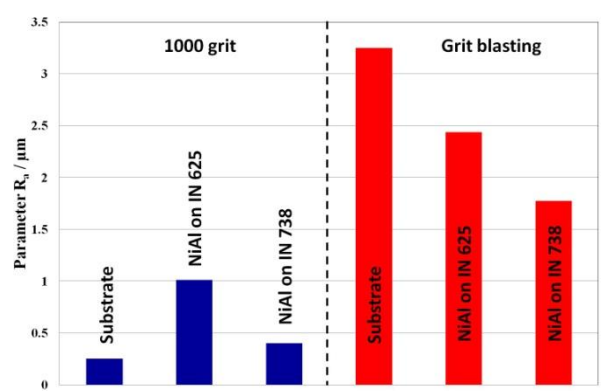

**Figure 3.** Comparison of roughness parameter $R_a$ measured on ground and grit-blasted substrates with aluminide coatings produced on these substrates.

The microstructures of the aluminide coatings manufactured on both Ni-based alloys with ground and grit-blasted surfaces are shown in Figure 4a–d. In all cross-sections, three zones can be clearly distinguished: An outer NiAl coating, the interdiffusion zone (IDZ), and substrates. In the cross-sections of the aluminide layers formed on grit-blasted IN 625 (Figure 4b) and IN 738 (Figure 4d), particles of $Al_2O_3$ present at the interface between the outer NiAl layer and the IDZ can be observed. These particles were the remaining particles used for the grit-blasting of the substrates that were built up into the near-surface region of the substrates. Therefore, they could be treated as local markers of the original surface which were overgrown by the NiAl layer. This observation confirms an outward mechanism of NiAl layer growth. The results of the layer thicknesses measurement are summarized in Table 2. For grit-blasted IN 625, a formation of a thicker outer NiAl layer was observed (21 μm) as compared with that formed on the ground alloy (17 μm). Regarding the cross-sections of aluminide layers formed on IN 625, a continuous and bright sub-layer in the IDZ was present in both the ground and grit-blasted surfaces. The thicknesses of the IDZ were comparable for both surface conditions. However, in the case of the grit-blasted substrate, this layer was more convoluted as compared to that on the ground surface. The surface preparation method of IN 738 did not affect the thickness of the outer NiAl layer. On the contrary, it strongly indicated that the thickness of the IDZ, namely the thickness of the IDZ obtained on grit-blasted IN 738 (18.93 μm) was almost two times higher than that observed on the ground substrate (10.90 μm). Moreover, a clear difference in the IDZ could be observed between the coatings manufactured on IN 625 and IN738. In the case of IN 625, a continuous layer in the IDZ could be observed, while for coatings on IN 738 a separate precipitates in the IDZ were present. This was probably caused by different alloy chemistry—namely, a higher amount of Ta and C in IN 738. The higher amount of Ta also influenced the microstructure of the alloy itself. Ta is known as a γ′-stabilizer which could lead to the presence of a γ-γ′ microstructure in the case of IN 738. Such a microstructure was not observed for IN 625. Moreover, a combination of the Ta presence and a higher concentration of C in the alloy led to the formation of carbides containing elements like Ti, Nb, Mo, and Ta in the case of IN 738, as observed previously [26].

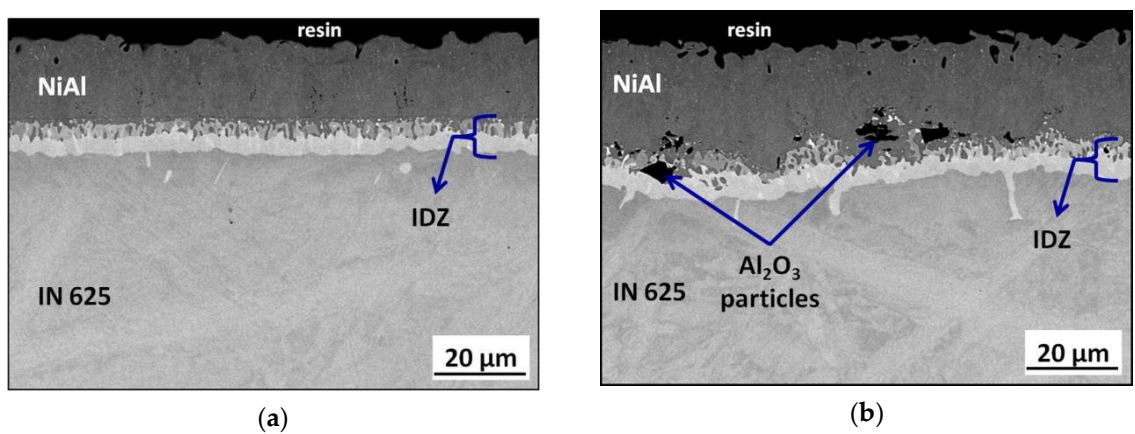

(**a**)          (**b**)

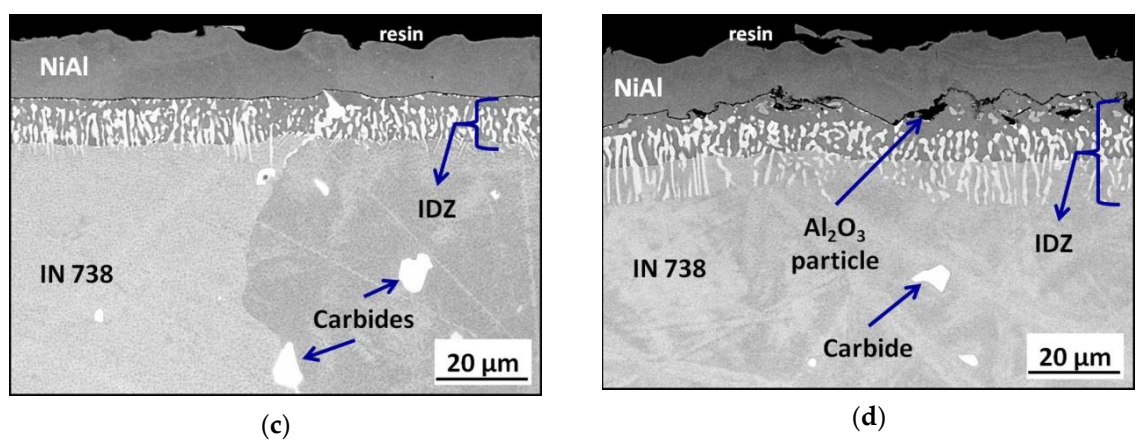

**Figure 4.** SEM/ back scattered electron (BSE) images showing cross-sections of aluminide coatings formed on: (**a**) Ground IN 625, (**b**) grit-blasted IN 625, (**c**) ground IN 738, and (**d**) grit-blasted IN 738 in the as-received condition.

**Table 2.** Measurements of the thickness of the outer NiAl layer and interdiffusion zone (IDZ) formed on ground and grit-blasted IN 625 and IN 738.

| Alloy | IN 625 | | IN 738 | |
|---|---|---|---|---|
| **Surface Preparation** | **1000 Grit** | **Grit-Blasting** | **1000 Grit** | **Grit-Blasting** |
| Outer NiAl thickness [μm] | 16.93 | 20.90 | 12.97 | 13.50 |
| Standard deviation | 1.36 | 1.77 | 1.77 | 2.06 |
| IDZ thickness [μm] | 8.13 | 11.70 | 10.90 | 18.93 |
| Standard deviation | 0.90 | 3.40 | 1.12 | 3.51 |

Figures 5 and 6 show the SEM/EDS elemental mappings obtained for aluminized IN 625 and IN 738, respectively. From these figures, it can be seen that outer coatings consisted of Ni and Al. The chemical analysis of the Ni and Al content in the outer NiAl coating performed by SEM/EDS revealed that the coating contained about 54 at.% of Ni and 38 at.% of Al. In the IDZ, an enrichment of Cr, Nb, and Mo was observed. The SEM/EDS elemental mapping obtained on the aluminide layer formed on ground IN 738 (Figure 6) revealed similar results, namely that the outer layer consisted of Ni and Al. Moreover, the SEM/EDS measurement of Ni and Al content revealed similar Ni and Al content, as in the case of the aluminide layer on IN 625. In the IDZ, a clear enrichment of chromium was observed. In addition, titanium and niobium enrichment could be found. However this enrichment probably indicates the place where Ti/Nb-carbide was formed in the substrate material. The GD-OES depth profiles obtained for aluminized IN 625 with the ground substrate are shown in Figure 7a. A depth profile revealed that the outer NiAl was non-stoichiometric, i.e., it contained more Ni than Al. This observation is in good agreement with measurement of the chemical composition of an NiAl coating using SEM/EDS. In the IDZ, a co-enrichment of carbon, chromium, molybdenum, and niobium was observed. This observation is in good agreement with findings by SEM/EDS maps. The GD-OES depth profile obtained for an aluminide coating on the grit-blasted substrate (Figure 7b) qualitatively showed a similar elements distribution. The main difference was in the profile measured for Al. Namely, in the IDZ, the concentration did not drop as sharply as in the case of the ground substrate—it decreased slowly. The latter was caused by the presence of $Al_2O_3$ particles originating from the grit-blasting of the surface. The GD-OES depth profile of the aluminized IN 738 with the ground surface (Figure 8a) showed a similar behavior as for IN 625; namely, the outer part of the NiAl coating contained more Ni than Al. In the IDZ, a co-enrichment of carbon, boron, and chromium was observed. Additionally, the co-enrichment of molybdenum and tungsten could be found. This indicates that the IDZ consisted of carbides and/or borocarbides of Cr, Mo, or W. The GD-OES depth profile for an aluminide coating on grit-blasted IN 738 (Figure 8b) revealed a similar trend in the profile for Al as found in the aluminide coating formed on grit-blasted

IN 625 due to the presence of $Al_2O_3$ particles. Qualitatively, the microstructures and distribution of the elements within the coating and the IDZ on the aluminide coating formed on IN 625 and IN 738 with the grit-blasted surface were similar to the respective coatings produced on ground substrates. Therefore, they are not shown here. To identify the phases present in the studied aluminide system, XRD analysis was performed. Figures 9 and 10 show the obtained XRD patterns for aluminized IN 625 and IN 738, respectively. The analysis confirmed findings by SEM and GD-OES, namely that the non-stoichiometric $Al0.42Ni0.58$ phase was detected as a major phase in the aluminide layers of all investigated samples. Additionally, in the systems based on IN 625, MoC and $Cr_{23}C_6$ phases were identified; for systems based on IN 738, only a $Cr_{23}C_6$ phase was found.

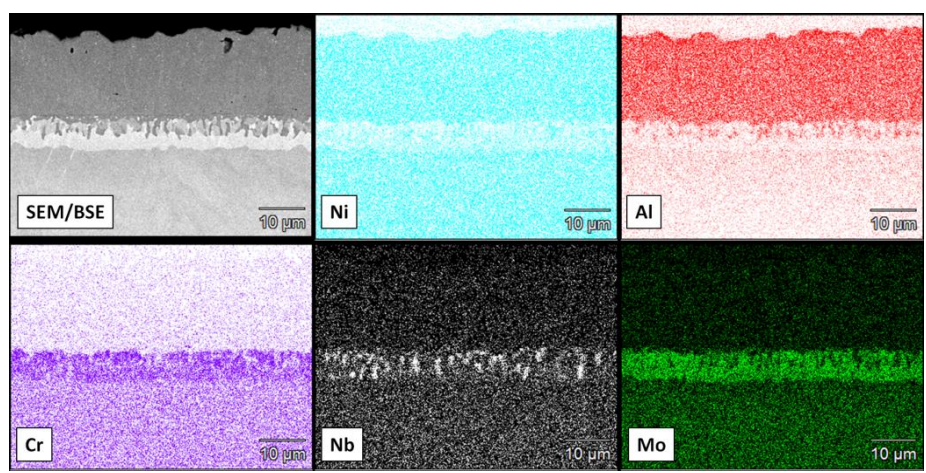

**Figure 5.** SEM/energy dispersive spectroscopy (EDS) elemental maps obtained on the cross section of aluminized IN 625 with the ground surface in the as-received condition. More intensive color indicates a higher concentration of given element.

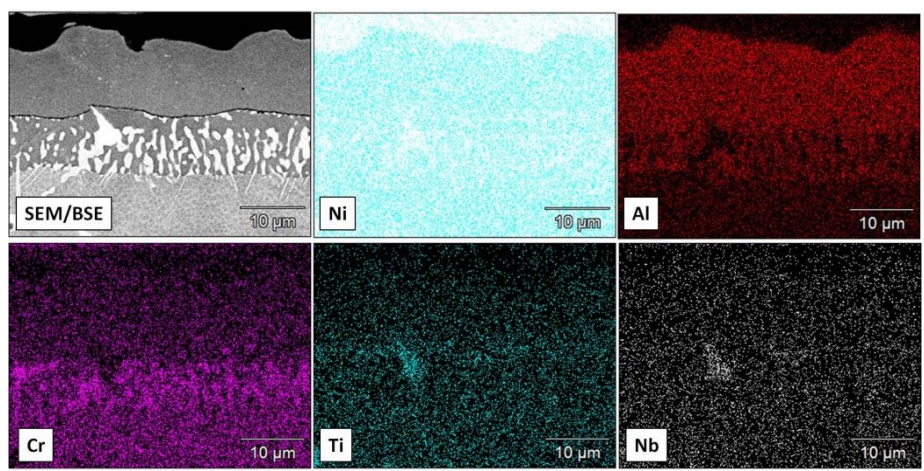

**Figure 6.** SEM/EDS elemental maps obtained on the cross section of aluminized IN 738 with the ground surface in the as-received condition. More intensive color indicates higher concentration of given element.

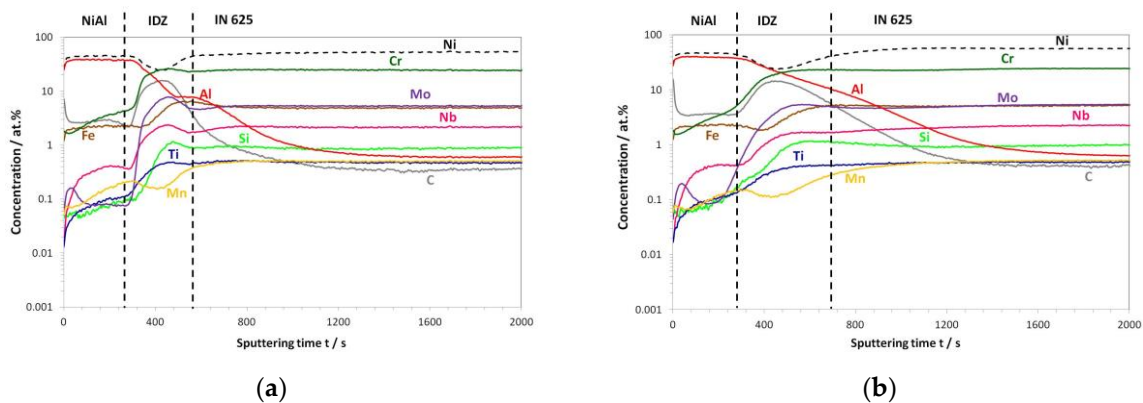

**Figure 7.** The glow discharge optical emission spectrometer (GD-OES) depth profiles obtained for the aluminide layer formed on: (**a**) Ground and (**b**) grit-blasted IN 625 in the as-received condition.

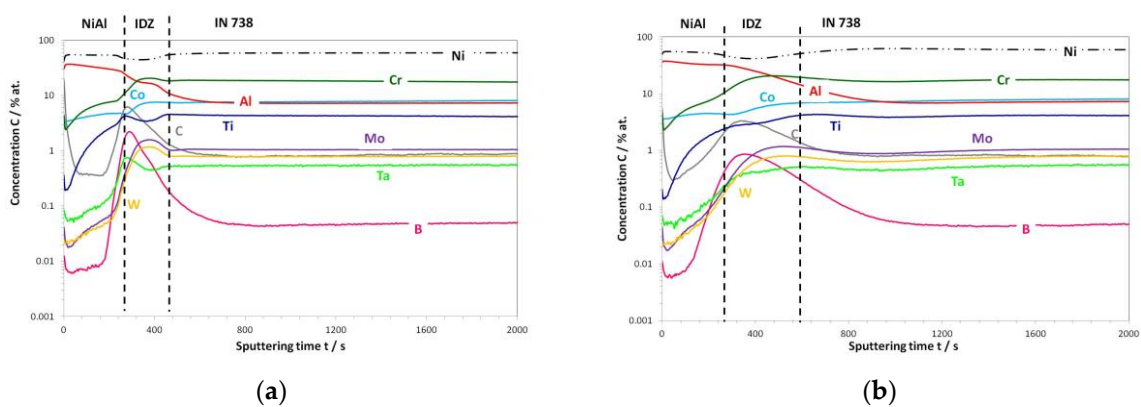

**Figure 8.** The GD-OES depth profile for aluminide layer formed on: (**a**) Ground and (**b**) grit-blasted IN 738 in the as-received condition.

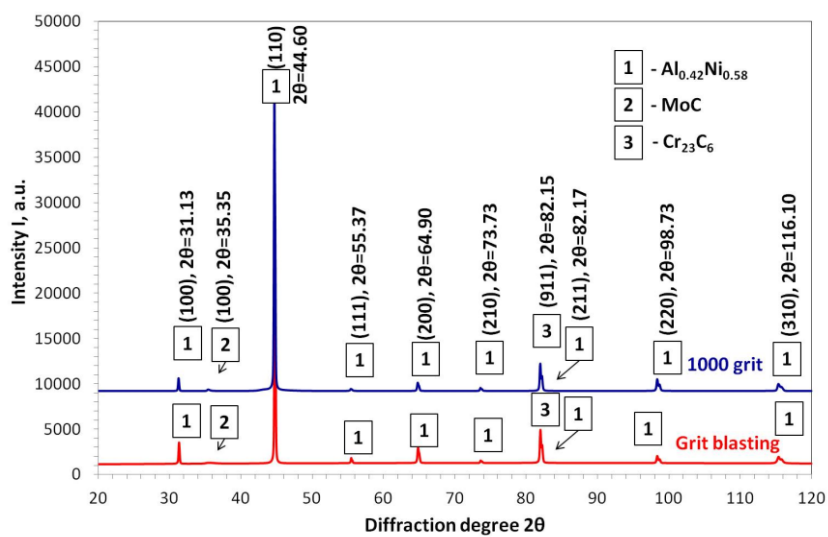

**Figure 9.** XRD patterns obtained from aluminized IN 625 with ground and grit-blasted surfaces in the as-received condition.

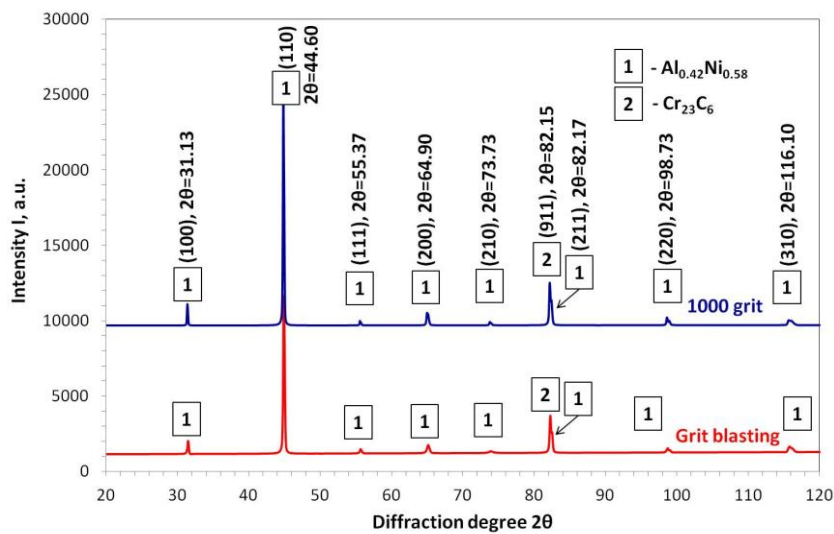

**Figure 10.** XRD patterns obtained from aluminized IN 738 with ground and grit-blasted surfaces in the as-received condition.

### 3.3. Post-Exposure Analyses

To elucidate the durability of the investigated aluminide coatings at an elevated temperature, a thermal shock test at 1120 °C in laboratory air was performed. The mass change curves obtained for the investigated aluminide systems are illustrated in Figure 11. As shown in the enlarged fragment of the plot, during the early stages of oxidation, all investigated samples revealed an increase of mass. The lowest mass gain appeared in aluminized IN 738 with the ground surface and appeared slightly higher in aluminized IN 738 with the grit-blasted surface. Finally, the highest mass gain was observed for both aluminized IN 625 (ground and grit-blasted) (see enlarged fragment of Figure 11). Considering the fact that the mass change was measured only on one sample per each preparation method, the differences in the mass change at the early stage of the oxidation could have been strongly influenced by the sample-to-sample variation. However, after the first 20 cycles, a slow mass loss was observed for all investigated systems. The higher drop in mass loss could be observed after 90 cycles (180 hot hours). The mass loss measured at end of the test (400 hot hours) was as follows: for ground and aluminized IN 738, $d_m = -1.28$ mg·cm²; for ground and aluminized, IN 625 $d_m$ = −11.72 mg·cm²; for grit-blasted and aluminized IN 738, $d_m$ = −16.99 mg·cm²; and, finally, for grit-blasted and aluminized IN 625 $d_m$ = −26.33 mg·cm². It is clear from the plot that aluminized IN 738 showed better oxidation resistance as compared to aluminized IN 625. Nevertheless, a higher mass loss was obtained for the grit-blasted and aluminized systems than for the ground and aluminized systems of both studied superalloys. Therefore, it is an obvious observation that grit-blasting decreased the durability of the aluminized systems at an elevated temperature. The decreasing mass change on Figure 11 indicates that the growth and spallation of formed oxide scale occurred. This led to a thinning of the aluminized coating and its depletion from Al. As observed in Figure 12, an enhanced spallation can be found for both the grit-blasted and aluminized alloys (see Figure 12b,d). The SEM/EDS analysis results show that all coatings formed aluminum rich oxide scales (Figures 12 and 13). The SEM/EDS chemical composition determination of an aluminide coating after exposure revealed that only about 4% was still present. Therefore, the nearly complete depletion of Al from the aluminide layer occurred. Moreover, for grit-blasted and aluminized IN 625, a complete spallation of an aluminide layer could be locally found, and, consequent, the formation of Ni/Cr-mixed oxide was observed (Figure 14). The latter resulted from the oxidation of a base alloy (IN 625). Additionally, a repeated cracking and alumina scale spallation led to the thinning of an aluminide layer, and the local discontinuity of an aluminide layer could be observed (Figure 15). This in turn, could lead to the diffusion cell effect described by Evans and Taylor [27] and cause massive oxidation. It can be suggested that foreign alumina particles coming from the grit-blasting process can locally suppress the diffusion of the elements at an elevated temperature, thus causing local differences in chemical composition. Moreover, due to the different coefficients of thermal expansion (CTE) of $Al_2O_3$ and the metallic substrates and coatings, an additional stress caused by the mismatch of CTEs of ceramic and metallic materials could be introduced during thermal cycling. This, in turn, could participate in the lower oxidation resistance of grit-blasted and aluminized systems. Additionally, no precipitates at the IDZ could be found after exposure—as was observed in the as-coated stage—which means that the IDZ dissolved due to the diffusion processes at the high temperature. Due to the severe degradation of aluminide coatings on IN 625, GD-OES depth profiling was not possible. Therefore, only the depth profiles after exposure of aluminide coatings produced on the ground (Figure 16a) and grit-blasted (Figure 16b) IN 738 are shown. The obtained results revealed the enrichment of oxygen and aluminum in the outer part of the scale, which indicates the formation of thermally grown $Al_2O_3$. The overall time for the measurement of the oxide scale region was larger for sample with the ground substrate surface (Figure 16a) as compared to the grit-blasted substrate surface (Figure 16b). This was probably caused by an enhanced oxide scale spallation from the aluminide coating on the grit-blasted substrate, as observed in the SEM images of the surfaces after exposure (Figure 12c,d). Similarly to the observations on the aluminide coatings in the as-received stage (Figure 8a,b), a region with the presence of embedded $Al_2O_3$ particles could be observed after exposure (Figure 16b). The presence of a plateau in the profiles measured for Al and O between 400 and 500 s of measurement supports the latter observation.

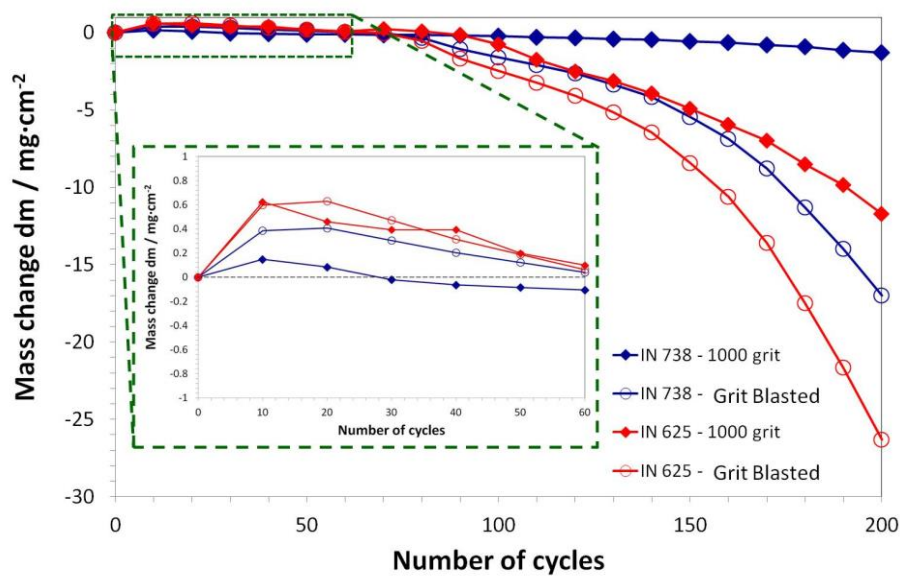

**Figure 11.** Mass changes plot obtained during thermal shocking test with cycles of 2 h heating and 15 min cooling, with pressurized air at 1120 °C in the air measured on aluminide coatings grown on IN 625 (red curves) and IN 738 (blue curves) with ground (fulfilled rectangular marks) and grit-blasted (circular open marks) surfaces.

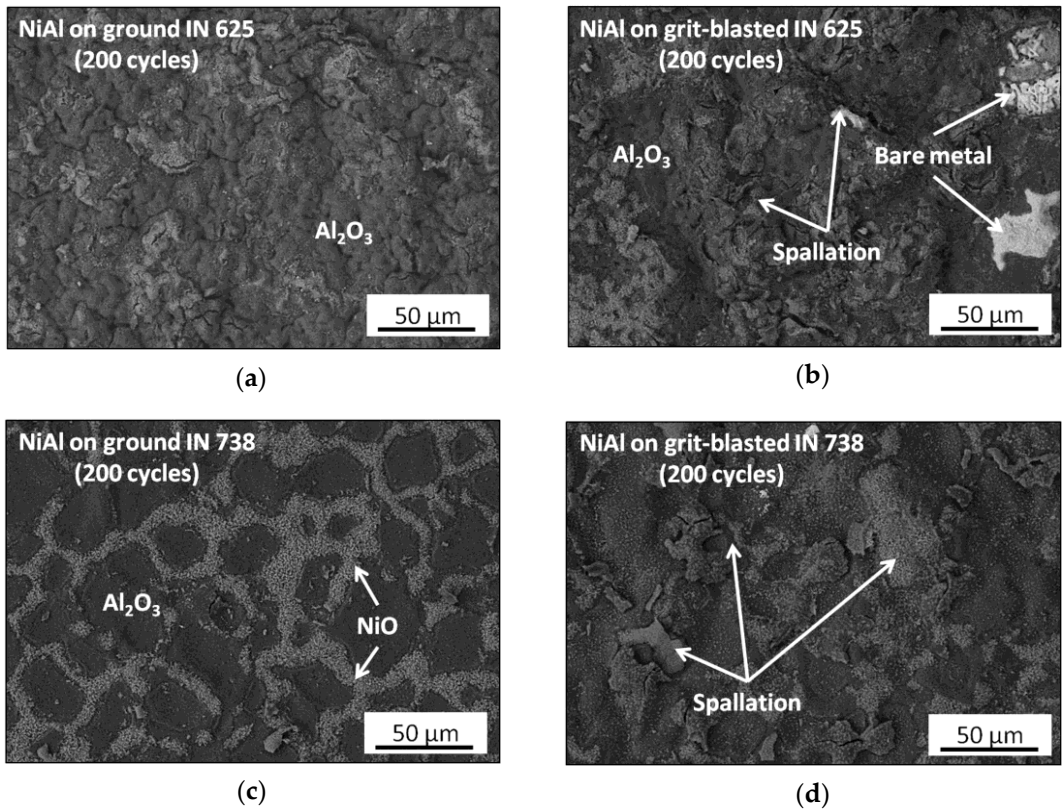

**Figure 12.** SEM/BSE images of the oxide scale formed on the surfaces of aluminide coatings grown on (**a**) ground IN 625, (**b**) grit-blasted IN 625, (**c**) ground IN 738, and (**d**) grit-blasted IN 738 after a thermal shocking test after 200 cycles.

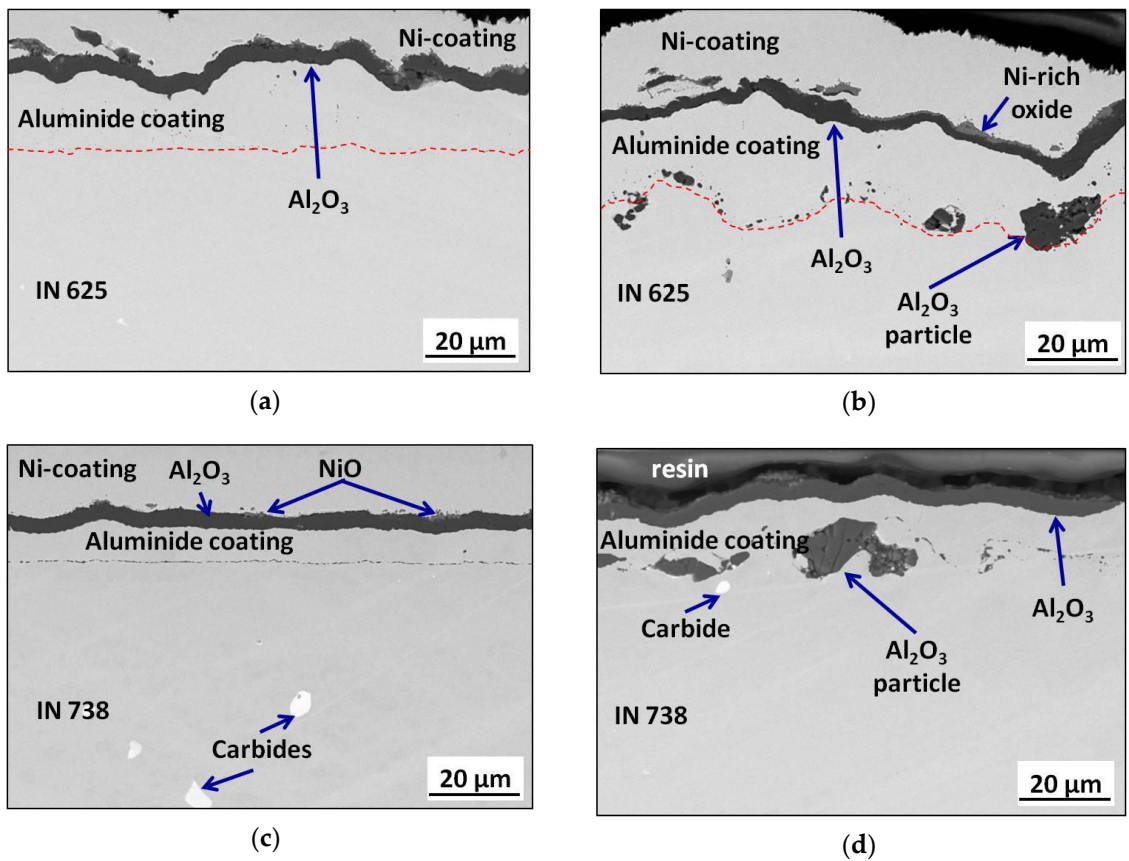

**Figure 13.** SEM/BSE images showing cross-sections of aluminide coatings formed on: (**a**) Ground IN 625, (**b**) grit-blasted IN 625, (**c**) ground IN 738, and (**d**) grit-blasted IN 738 after a thermal shocking test after 200 cycles.

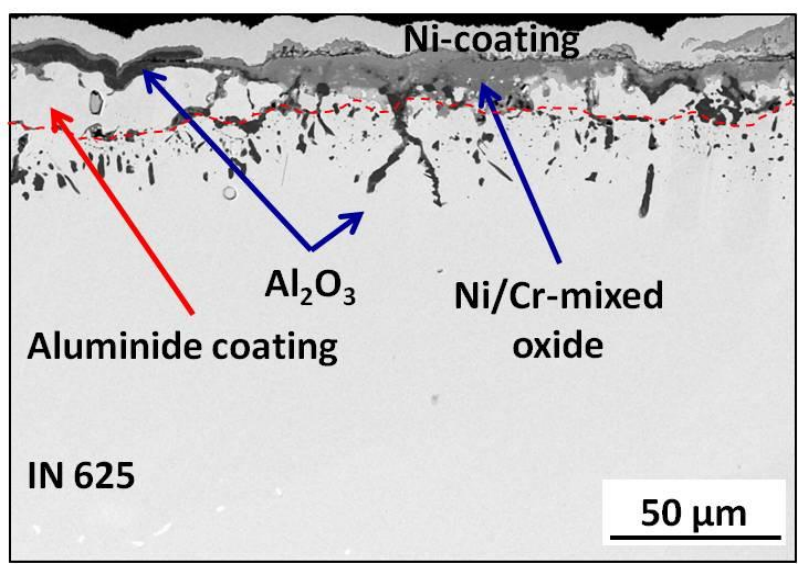

**Figure 14.** SEM/BSE image showing cross-section of an aluminide coating formed on grit-blasted IN 625 after a thermal shocking test with after 200 cycles, showing the area with completely spalled-off aluminide coating.

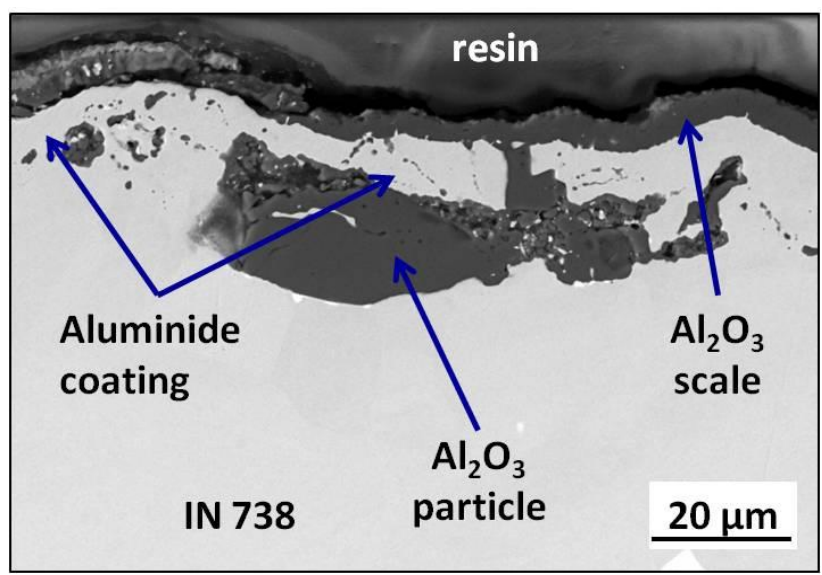

**Figure 15.** SEM/BSE image showing cross-section of an aluminide coating formed on grit-blasted IN 738 after a thermal shocking test after 200 cycles, showing the influence of embedded alumina particle originating from grit blasting of the substrate surface on the oxidation behavior of the complete system.

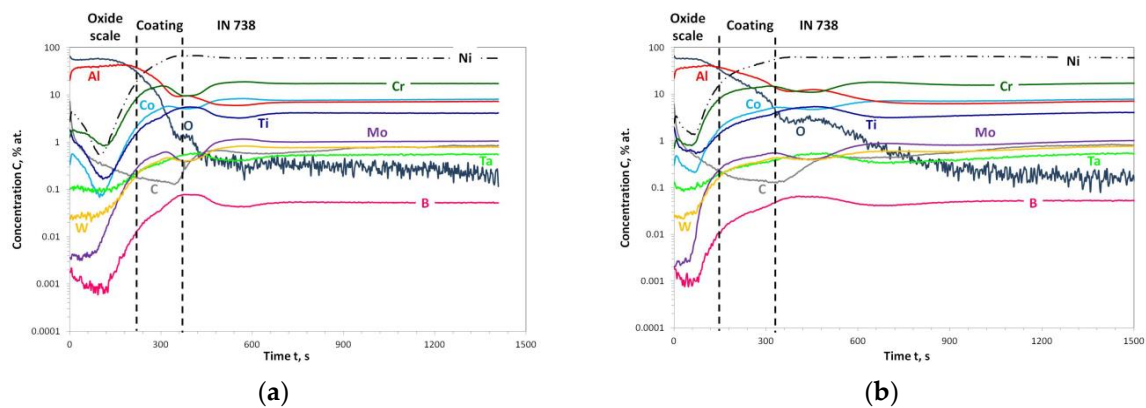

**Figure 16.** The GD-OES depth profile obtained for aluminide layer formed on: (**a**) Ground and (**b**) grit-blasted IN 738 after 200 cycles of a thermal shocking test.

## 4. Conclusions

Base on the results obtained in the present work, the following conclusions can be made:

- Surface mechanical treatment influences the thickness and morphology of aluminized Ni-based superalloys: for grit-blasted IN 625, the thickening of NiAl coating was observed. Tor grit-blasted IN 738, a thicker IDZ was found.
- The presence of foreign particles, namely alumina, coming from grit-blasting process negatively affects diffusion processes at an elevated temperature. In addition, they can introduce additional stresses during the heating and cooling of the samples at the coating/substrate interface.
- The grit-blasting of substrates results in a worse oxidation behavior of studied aluminized systems.

**Author Contributions:** Conceptualization, W.J.N. and B.W.; data curation, W.J.N., P.W., K.G. and B.W.; formal analysis, W.J.N. and B.W.; funding acquisition, W.J.N.; investigation, W.J.N., K.O., P.W. and K.G.; resources, W.J.N. and K.O.; supervision, B.W.; validation, methodology, project administration, writing—original draft, and writing—review & editing, W.J.N.

**Funding:** This research was financed within the Marie Curie COFUND scheme and POLONEZ program from the National Science Centre, Poland. POLONEZ grant No. 2015/19/P/ST8/03995. This project has received funding from the European Union's Horizon 2020 research and innovation programme under the Marie Skłodowska-Curie grant agreement No. 665778.